\documentclass[prl,twocolumn,superscriptaddress,longbibliography]{revtex4-1}
\usepackage{amssymb,amsmath,amsfonts,bm}
\usepackage{graphicx}
\usepackage{epstopdf}
\usepackage{times}
\usepackage{float}
\usepackage{lipsum}
\usepackage{color}

\usepackage{hyperref}
\hypersetup{  colorlinks=true, linkcolor=blue, citecolor=red, urlcolor=blue  }

\usepackage{physics}

\begin{document}

\title{Brachistochrone Non-Adiabatic Holonomic Quantum Control}
%

\author{Bao-Jie Liu}

\affiliation{Department of Physics, Southern University of Science and Technology, Shenzhen 518055, China}

\author{Zheng-Yuan Xue} \email{zyxue83@163.com}
\affiliation{Guangdong Provincial Key Laboratory of Quantum Engineering and Quantum Materials, 
and School of Physics\\ and Telecommunication Engineering, South China Normal University, Guangzhou 510006, China}

\affiliation{Guangdong-Hong Kong Joint Laboratory of Quantum Matter, and Frontier Research Institute for Physics, South China Normal University, Guangzhou 510006, China}

\author{Man-Hong Yung}  \email{yung@sustech.edu.cn}
\affiliation{Department of Physics, Southern University of Science and Technology, Shenzhen 518055, China}
\affiliation{Shenzhen Institute for Quantum Science and Engineering, Southern University of Science and Technology, Shenzhen 518055, China}
\affiliation{Guangdong Provincial Key Laboratory of Quantum Science and Engineering, Southern University of Science and Technology, Shenzhen 518055, China}
\affiliation{Shenzhen Key Laboratory of Quantum Science and Engineering, Southern University of Science and Technology, Shenzhen,518055, China}

\date{\today}

\begin{abstract}
In quantum control, geometrical operations could provide an extra layer of robustness against control errors. However, the conventional non-adiabatic holonomic quantum computation (NHQC) is limited by the fact that all of the operations require exactly the same amount of evolution time, even for a small-angle rotation. Furthermore, NHQC confines the driving part of the Hamiltonian to strictly cover a fixed pulse area, making it sensitive to control errors. Here we present an unconventional approach of NHQC, termed B-NHQC, for bypassing these limitations. Specifically, with B-NHQC, non-Abelian geometric gates can be time-optimized by following the brachistochrone curve, minimizing the impact from the environmental decoherence. Additionally, we demonstrate that B-NHQC is compatible with composed pulses, which can further enhance the robustness against pulse errors. For benchmarking, we provide a thorough analysis on the performance of B-NHQC under experimental conditions; we found that the gate error can be reduced by as much as 64\% compared with NHQC.
\end{abstract}


\maketitle
\emph{Introduction.}---Realizing precise and noise-resistant quantum gates are of vital importance to the success of quantum computation. Geometric phases, depending only on the global properties of the evolution paths, are immune to local errors~\cite{Zanardi1999,gqc,b1}. Therefore, geometric quantum computation ~\cite{Jones,Zhu2005,Berger2013,Chiara,Leek,Filipp}, where quantum gates are induced by geometric phases, is a promising strategy for fault-tolerant quantum computation. In fact, a geometric phase can either be a real number (Abelian), also known as ``Berry's phase"~\cite{b2},  or a matrix (non-Abelian)~\cite{b3}, which is the key ingredient in constructing quantum operations for holonomic quantum computation.

Holonomic quantum computation based on adiabatic evolution has been proposed~\cite{Duan2001a,lian2005}. However, the adiabatic condition implies a lengthy gate time, thus the environment-induced decoherence would introduce severe gate infidelity. To overcome this problem, nonadiabatic holonomic quantum computation (NHQC) have been proposed to remove the adiabatic condition
~\cite{Sjoqvist2012,Xu2012}. Moreover, NHQC can be extended to realize arbitrary single-qubit holonomic quantum gate in a single-loop/shot evolution
\cite{Xu2015,E2016,Herterich2016,Zhao2017,Hong2018} and compatible with elementary optimal control techniques \cite{liuprl,lisaiaqt,lisaiexp}. Recently,  elementary holonomic quantum gates have been experimentally demonstrated in different platforms, including superconducting circuits~\cite{Abdumalikov2013,Egger2019,xu2018,Yan2019}, nuclear magnetic resonance (NMR) \cite{Feng2013,li2017,zhu2019}, and nitrogen-vacancy centers in diamond~\cite{,zu2014,Arroyo2014,Sekiguchi2017,Zhou2017,N2018,K2018}, etc. However, these NHQC implementations are sensitive to systematic errors due to the stringent requirements on the govern Hamiltonian~\cite{Thomas,Johansson,Zheng,jun2017,Ramberg2019}.

Here, we present an unconventional approach of NHQC, called brachistochrone NHQC (B-NHQC), for breaking the two limitations of the conventional NHQC, which extends the unconventional Abelian geometric scheme~\cite{Zhu2003,Du2006} the non-Abelian case.  Comparing with previous implementations of conventional NHQC, our scheme  is more robust against control induced imperfections, as the fixed pulse area limitation is removed.  Moreover, we can combine our model with various composite-pulse schemes to further enhance the robustness of B-NQHC. Furthermore, our scheme also incorporates with the time-optimal technology~\cite{Alberto2007, Carlini2008, Carlini2012, Carlini2013, Wang2015, Du2017}, by solving the quantum brachistochrone equation (QBE), to realize universal holonomic gates with \emph{minimum} time, and thus minimizes the decoherence induced gate infidelity. For demonstration, we consider a three-level quantum system to explain the working mechanism of B-NHQC. Numerical simulations indicate that our B-NHQC and its' optimization can achieve a significant improvement over the NHQC gates~\cite{Sjoqvist2012, Xu2012} using experimental parameters. Note that our work is different from the time-saving single-shot NHQC~\cite{XGF2018} with off-resonance drives, which is basically incompatible with pulse shaping and experimentally difficult.

\emph{General model.}---We consider a general time-dependent Hamiltonian~$H(t)$. Assume $\left\{ {\left| {{\psi _k}\left( t \right)} \right\rangle } \right\}$ is a complete set of basis, following the time-dependent Schr\"{o}dinger equation (TDSE), i.e.,  $i|\dot{\psi}_{k}(t)\rangle= H(t)\left|\psi_{k}(t)\right\rangle$. The time-evolution operator is given by $U\left( t,0 \right) = {\mathcal T}{e^{ - i\int_0^t {H\left( {t'} \right)} dt'}} = \sum\nolimits_m {\left| {{\psi _m}\left( t \right)} \right\rangle \left\langle {{\psi _m}\left( 0 \right)} \right|}$. In addition, we introduce a different set of time-dependent basis, $\left\{ {\left| {{\phi _k}\left( t \right)} \right\rangle } \right\}$, satisfying the boundary conditions at time $t=0$ and $t=\tau$, namely $
\left| {{\phi _k}\left( \tau  \right)} \right\rangle  = \left|{ {\phi _k}\left( 0 \right)} \right\rangle=\left|{ {\psi _k}\left( 0 \right)} \right\rangle$. Substituting  the solutions into TDSE and applying the boundary conditions, the unitary transformation matrix at the final time $t=\tau$ can be expressed~\cite{liuprl} as $U(\tau)=\sum_{ml}\left[\mathcal{T} \mathrm{e}^{\mathrm{i}\int_{0}^{\tau}\left(A+K\right)dt}\right]_{ml} \left|\psi_{m}(0)\right\rangle\left\langle\psi_{l}(0)\right|$, where $\mathcal{T}$ is time ordering operator,  the dynamical and geometric parts are denoted by ${K_{ml}} \equiv-\left\langle {{\phi _m}\left( t \right)} \right|H\left( t \right)\left| {{\phi _l}\left( t \right)} \right\rangle$ and ${A_{ml}} \equiv i\left\langle {{\phi _m}\left( t \right)} \right|\frac{d}{{dt}}\left| {{\phi _l}\left( t \right)} \right\rangle$.

Here, we first extend the definition of the unconventional geometric phase~\cite{Zhu2003,Du2006} from the Abelian case to a non-Abelian case. Specifically, our strategy is to confine the auxiliary basis $\left\{ {\left| {{\phi _k}\left[\lambda_{a}(t),\eta_{b}(t) \right]} \right\rangle } \right\}(a,b=1,2,...,n)$ with two sets of independent parameters $\lambda_{a}(t)$ and $\eta_{b}(t)$. Consequently, the geometric part $A_{ml}(t)$ becomes $A_{m l}=A_{m l}^{\lambda}+A_{m}^{\eta}$ where $A_{m l}^{\lambda}\equiv\sum_{a} i\langle\phi_{m}|\frac{\partial}{\partial \lambda_{a}}| \phi_{l}\rangle\left(d \lambda_{a} / d t\right)$ and $A_{m l}^{\eta}\equiv\sum_{b} i\langle\phi_{m}|\frac{\partial}{\partial \eta_{b}}| \phi_{l}\rangle\left(d \eta_{b} / d t\right)$. To make the evolution operator $U(\tau)$ purely geometric, the dynamical parts $K_{ml}$ is offset by the geometric part $A^{\eta}_{ml}$ at each moment, i.e.,
\begin{equation}\label{UNHQC}
    K_{m l}+A_{m l}^{\eta}=0 \ .
\end{equation}
Consequently, see Appendix A for detail proof, Eq.~(\ref{UNHQC}) implies that the unitary operator becomes \emph{purely} geometrical
\begin{equation}\label{UNHQC3}
U(\tau)=\mathcal{P}\mathrm{e}^{\mathrm{i}\oint_{0}^{\tau}A_{a}^{\lambda}d\lambda_{a}} \ .
\end{equation}
where $\mathcal{P}$ is path  ordering along the closed path.

\begin{figure}[tb]
\includegraphics[width=7cm]{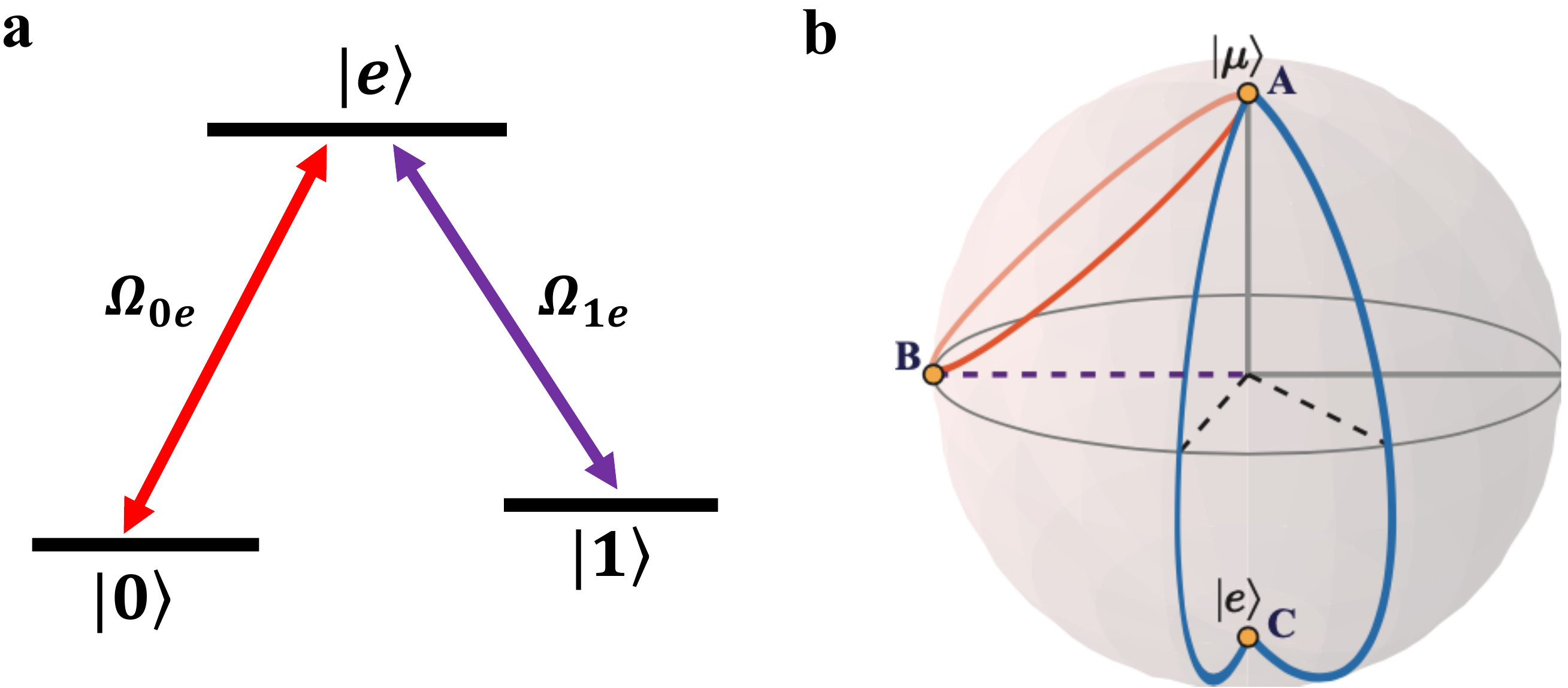}
\caption{ Illustration of the proposed scheme. (a) The coupling configuration with two microwave fields $\Omega_{0e}$ and $\Omega_{1e}$ resonantly coupled to the three levels of a quantum system. (b) Geometric illustration of the proposed B-NHQC gate on the Bloch sphere, where the state $|\mu\rangle$ undergoes a cyclic evolution A$\rightarrow$B$\rightarrow$A. However, the conventional NHQC takes a cyclic evolution A$\rightarrow$C$\rightarrow$A.}
\label{fig1}
\end{figure}

A possible choice satisfying the condition in Eq.~(\ref{UNHQC}) would be $ {\left| {{\phi_a}\left[\lambda_{a} (t),\eta_{b}(t) \right]} \right\rangle }=e^{i\lambda_{a}(t)}|\psi_{a}\left[\eta_{b}(t)\right]\rangle$, where $|\psi_{a}\left[\eta_{b}(t)\right]\rangle$ evolves according to the TDSE. Together with the boundary condition, the unconventional holonomic gate in Eq. (\ref{UNHQC3}) becomes diagonal:
\begin{equation}\label{UHOLONOMY}
U(\tau)=\sum_{a} e^{i \lambda_{a}(\tau)}\left|\phi_{a}(0)\right\rangle\left\langle\phi_{a}(0)\right| \ ,
\end{equation}
where  geometric phase $\lambda_{a}=i \oint\left\langle\phi_{a}\left[\lambda_{a}(t), \eta_{b}(t)\right]\left|\partial_{\eta_{b}}\right| \phi_{a}\left[\lambda_{a}(t), \eta_{b}(t)\right]\right\rangle d \eta_{b}$ is  adjustable via the parameter $\eta_{b}(t)$

Note that similar gate form can also be obtained in the  conventional NHQC schemes~\cite{Sjoqvist2012,Xu2012} but with  the constraint of $K=0$ to ensure the gate to be holonomic.  However, our B-NHQC scheme   removes this constraint by using unconventional holonomy, which makes it possible to incorporate our scheme with optimal control technology to suppress the effect of different errors, as discussed below. 
Although $\lambda_{a}$ is an Abelian A-A phase~\cite{b1}, geometric part $A$ does represent a non-abelian connection with non-vanishing commutation relation $[ A(t), A(t')]\neq 0$, where proves the non-Abelian nature of the gate~\cite{E2016} similar to conventional NHQC cases~\cite{Sjoqvist2012,Xu2012,E2016}. In addition, we emphasize that the results is independent on the choice of the auxiliary states $\left\{ {\left| {{\phi _k}\left[\lambda_{a}(t),\eta_{b}(t) \right]} \right\rangle } \right\}$, as they are generally chosen.

Inevitably, any implementation will suffer from decoherence, which reduces the target gate fidelity, and thus operations with \emph{minimum} time becomes a preferable choice for realizing high-fidelity gates. To realize the fastest geometric gates, we extend the above framework to B-NHQC, by combining it with the time-optimal control technique~ \cite{Alberto2007,Carlini2008,Carlini2012,Carlini2013,Wang2015,Du2017}, via solving the QBE
\begin{equation}\label{QBE}
i  \partial F /\partial t  =[H,F],
\end{equation}
where  $F=\partial { L }_{ c }/\partial { H }$ and ${ L }_{ c }={ \sum   }_{ j }{ \mu  }_{ j }{ f }_{ j }\left( { H } \right)$, with ${ \mu }_{ j }$ being the Lagrange multiplier. Note that, choosing different  parameters of the driven Hamiltonian $H(t)$ makes the evolution of the system follow different paths, but leads to a same unconventional holonomic gate. The path with the minimal time can be obtained by solving the QBE together with TDSE. For a realistic physical systems, there always be a constraint that the energy bandwidth, described by ${ f }_{ 0 }\left( H \right) =\left[ Tr\left( { H }^{ 2 } \right) -{ E }^{ 2 } \right] /2$, should be finite.

\emph{Application of B-NHQC.}---We firstly illustrate the implementation of our idea in a three-level system, where the ground state $|0\rangle$ and $|1\rangle$ are chosen as logic states of the qubit, while the excited state $|e\rangle$ as an auxiliary state, as shown in Fig.~\ref{fig1}(a). The transitions of $|0\rangle \leftrightarrow |e\rangle$ and $|e\rangle \leftrightarrow |1\rangle$ are driven resonantly by two microwave fields, with the amplitudes $\Omega_{0e}(t)$, $\Omega_{e1}(t)$ and the phases $\phi_{0}$  and $\phi(t)$, respectively. Assuming $\hbar = 1$ hereafter, the Hamiltonian of the system is
\begin{eqnarray} \label{h1}
H_{1}(t) =\frac{\Omega(t)}{2} e^{-i\phi(t)} \left(\sin\frac{\theta}{2}e^{i\phi_{1}}|0\rangle +\cos\frac{\theta}{2}|1\rangle\right)\langle e| + \mathrm{H.c.},\notag\\
\end{eqnarray}
where $\phi_{1}=\phi_0-\phi(t)$, $\Omega(t)=\sqrt{\Omega_{0e}^2(t)+\Omega_{1e}^2(t)}$ and  $\tan(\theta/2)=\Omega_{0e}(t)/\Omega_{1e}(t)$, and  the mixing angle $\theta$ and phase $\phi_{1}$ are set to be time-independent.

Recall the construction of our unconventional holonomy, we need to choose a set of auxiliary states, satisfying cyclic condition $|\phi_{m}(0)\rangle=|\phi_{m}\left[\lambda_{a}(\tau),\eta_{b}(\tau) \right]\rangle$. Here we choose $|\phi_{0}\rangle=e^{-i \lambda_{1}}|\psi_{0}{(\eta_{b})}\rangle$, $\left|\phi_{1}\right\rangle=e^{i \lambda_{1}}|\psi_{1}{(\eta_{b})}\rangle$ and $\left|\phi_{2}\right\rangle=|\psi_{2}\rangle$ with
\begin{eqnarray}\label{ThreST}
\left|\psi_{0}\right\rangle &=&\left( c_{\frac{\eta_{1}}{2}}-i s_{\frac{\eta_{1}}{2}} c_{\eta_{3}}\right) e^{i \frac{\eta_{2}}{2}}|e\rangle-i s_{\eta_{3}} s_{\frac{\eta_{1}}{2}} e^{-i \frac{\eta_{2}}{2}}|\mu\rangle , \notag\\
\left|\psi_{1}\right\rangle &=& -i s_{\eta_{3}} s_{\frac{\eta_{1}}{2}} e^{i \frac{\eta_{2}}{2}}|e\rangle+\left( c_{\frac{\eta_{1}}{2}}+i s_{\frac{\eta_{1}}{2}} c_{\eta_{3}}\right) e^{-i \frac{\eta_{2}}{2}}|\mu\rangle , \notag\\
\left|\psi_{2}\right\rangle &=& c_{\frac{\theta}{2}}|0\rangle-s_{\frac{\theta}{2}} e^{-i \phi_{1}}|1\rangle \ ,
\end{eqnarray}
where we set $s_{x} \equiv \sin x$, $c_{x} \equiv \cos x$, $\dot{\eta}_{3}=0$ and $|\mu\rangle=s_{\frac{\theta}{2}}e^{i \phi_{1}}|0\rangle+c_{\frac{\theta}{2}} |1\rangle$ for simplicity.

By directly substituting Eq. (\ref{ThreST}) into TDSE, we found that the control parameters of microwave pulses are determined by 
\begin{equation}\label{TSD}
\Omega(t) = -\dot{\eta}_{1}s_{\eta_{3}}, \quad \phi(t)=\eta_{2}(t)=\eta_{1}/c_{\eta_{3}} \ .
\end{equation}
Using the Eq. (\ref{ThreST}), the dynamical parts and geometric parts can be obtained under the condition in Eq.~(\ref{UNHQC}).
When the cyclic evolution condition $\eta_{1}(\tau)=2\pi$ ($\left|\phi_{1}(0)\right\rangle =|\mu\rangle$) is met with the path A$\rightarrow$B$\rightarrow$A shown in Fig.~\ref{fig1}(b), the unconventional holonomy in Eq. (\ref{UNHQC3}), in computational subspace $\left\{ |\phi_{1}(0)\rangle, |\phi_{2}(0)\rangle \right\}$, is
 \begin{equation}\label{BUNHQC}
U(\tau)=e^{i\gamma}\left|\phi_{1}(0)\right\rangle\left\langle\phi_{1}(0)\right| +\left|\phi_{2}(0)\right\rangle\left\langle\phi_{2}(0)\right|,
 \end{equation}
where $\gamma=\lambda_{1}(\tau)-\lambda_{1}(0)=\pi-\eta_{2}(\tau)/2$ is a geometric phase.
Note that 
$U(\tau)$  can be spanned in the logical qubit subspace $\left\{|0\rangle,|1\rangle\right\}$ as
\begin{eqnarray}\label{UU}
{U}(\theta,\phi_{1},\gamma)
&=& \exp\left(i\frac{\gamma}{2}\right) \exp\left(  - i\frac{\gamma }{2}{\mathbf{n}} \cdot \sigma \right),
\end{eqnarray}
which describes a rotation around the $\textbf{n}=(\sin\theta\cos\phi,\sin\theta\sin\phi,\cos\theta)$ axis by a $\gamma$ angle. Since $\textbf{n}$ and $\gamma$ can be arbitrary, $U(\theta,\phi_{1},\gamma)$ can construct arbitrary single-qubit gates, in a single loop evolution. Moreover, our scheme can reduce to the conventional NHQC scheme~\cite{Sjoqvist2012,Xu2012}, simply by setting $\eta_{3}=\pi/2$ and $\eta_{2}=0$. Then, the evolution satisfies the parallel transport condition, and the unconventional geometric phase $\gamma=\pi$ becomes a pure geometric phase.

\begin{figure}[tb]
	\centering
   \includegraphics[width=8cm]{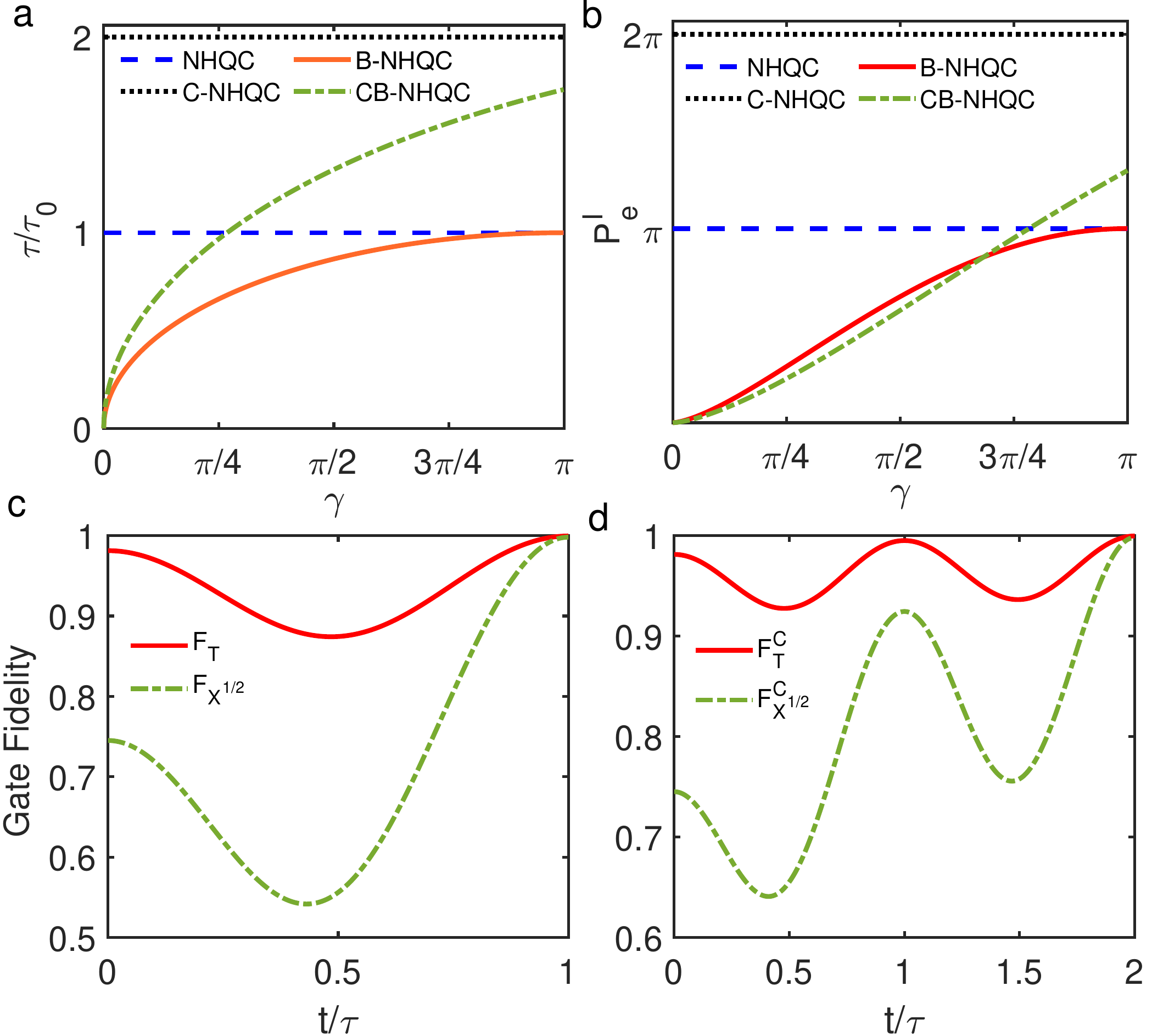}
\caption{\label{fig2}(a) The dimensionless gate time and (b) the integrated excited-state population $P_{e}^{I}$ of B-NHQC, CB-NHQC, NHQC and C-NHQC schemes with respect to the rotation angel $\gamma$. Dynamics of gate fidelities of  $X^{1/2}$ and \emph{T} gates in B-NHQC (c) and CB-NHQC (d) schemes.}
\end{figure}

Recall that for realizing B-NHQC gates, we need to minimize the above gate time by solving the QBE in Eq. (\ref{QBE}) to reduce the influence of the decoherence effect.
Here, we can simply set $\Omega(t)=\Omega_{0}$ to satisfy the constraint, i.e., ${ f }_{ 0 }\left( H_{1} \right) =\left[ Tr\left( { H_{1} }^{ 2 } \right) -{ \Omega_{0} }^{ 2 } \right] /2=0$. Following the Ref.~\cite{Wang2015,Du2017}, by solving the Eq.~(\ref{QBE}), one minimum-time solution to our purpose is $\phi \left( t \right) =\eta_{2} \left( t \right) =2(\gamma -\pi)t/\tau$, with the minimum evolution time being $\tau =2\sqrt { { \pi  }^{ 2 }-{ \left( \pi -\gamma  \right)  }^{ 2 } } /\Omega _{ 0 }$, which decreases with the decrease of the geometric phase $\gamma$.

Furthermore, we can enhance the robustness of B-NHQC against systematic control errors by combining it with the composite pulse strategy, which we call CB-NHQC, similar to the case of composite NHQC (C-NHQC)~\cite{XU2017,zhu2019}. To achieve this, the CB-NHQC is divided into two parts, i.e., $(0, \tau)$ and $(\tau, 2\tau)$.  During  the first interval $(0 \leq t \leq \tau)$, we set the phase
$\phi \left( t \right)=2(\gamma/2 -\pi)t/\tau$ corresponding to the evolution operator ${U_{1}}(\theta,\phi_{1},\gamma/2)$.
For the second interval,  $\phi' \left( t \right)=\pi+2(\gamma/2 -\pi)t/\tau$ with the evolution operator $U_{2}=-U_{1}$. As a result, the obtained CB-NHQC gate is
${U_{C}}(\theta,\phi_{1},\gamma)=-[{U_{1}}(\theta,\phi_{1},\gamma/2)]^2$.
Here, we plot the evolution time $\tau$, in unit of $\tau_0=2\pi/\Omega_{0}$, of B-NHQC, CB-NHQC, NHQC and C-NHQC as a function of their corresponding geometric phases, as shown in Fig. \ref{fig2}(a), which clearly shows that B-NHQC scheme generally has the shortest gate time.
Moreover, as shown in  Fig. \ref{fig2}(b), B-NHQC and CB-NHQC schemes can greatly reduce the integrated excited-state populations $\mathrm{P}_{\mathrm{e}}^{I}=\int_{0}^{\tau}|\langle\psi(t) | e\rangle|^{2} dt$ compared with NHQC and C-NHQC corresponding to the initial state $|\mu\rangle$, which are beneficial to reduce the excited-state decay.

The performance of our proposal in Eq. (\ref{UU}) can be evaluated by using the quantum master equation.
In our simulation, we choose the parameters from the the current experiment \cite{Buluta2011}, the decay and dephasing rates of all the transmons to be the same as 
$\Gamma=\Omega_{0}/2000$. We evaluate the \emph{T} gate $U_{T}=U(0,0,\pi/4)$ and the   $X^{1/2}$ gate $U_{X^{1/2}}=U(\pi/2,0,\pi/2)$, using the gate fidelity defined by $F_{ T,X^{1/2}  }=\frac { 1 }{ 2\pi  } \int _{ 0 }^{ 2\pi  }{ \left< \psi _{ T,X^{1/2} } |{ { \rho  }_{ 1 } }|{ { \psi _{ T,X^{1/2}  } } } \right>  } d{ \chi  }_{ 1 }$ for a general initial state of $|\psi(0)\rangle=\cos\chi_{1}|0\rangle+\sin\chi_{1}|1\rangle$ with the target state being $|{\psi _{ T,X^{1/2} } } \rangle=U_{T,X^{1/2}}|\psi\rangle$. As show in Figs. \ref{fig2}(c) and \ref{fig2}(d), we plot gate fidelities as functions of the time $\tau$ for 1001 input states with $\chi_{1}$ uniformly distributed over $[0,2\pi]$, and the obtained T gate and $X^{1/2}$ fidelities in B-NHQC are $F _{X^{1/2}}= 99.81\%$ and $F _{T}= 99.90\%$, and $F ^C _{X^{1/2}}= 99.82\%$ and $F _{T} ^C= 99.92\%$ in CB-NHQC.

\begin{figure}[tb]
	\centering
\includegraphics[width=8cm]{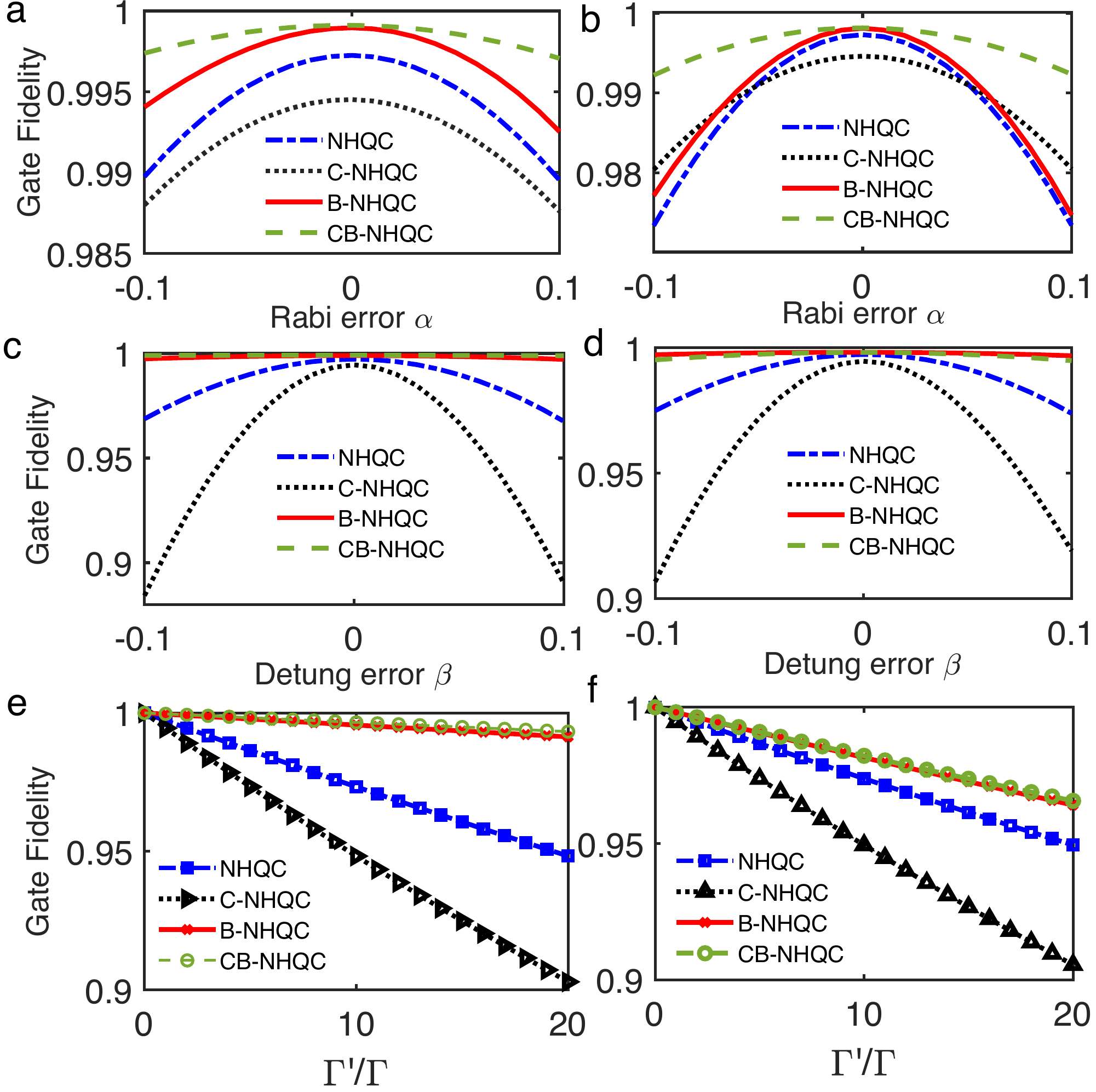}
\caption{\label{fig3} The gate fidelities under imperfections. The \emph{T} and $X^{1/2}$ gate fidelities for B-NHQC, CB-NHQC, NHQC and C-NHQC cases under the Rabi error (a) and (b), detuning error (c) and (d), decoherence (e) and (f), respectively.}
\end{figure}

We now proceed to show the robustness improvement of our scheme.  Firstly, to investigate the robustness against pulse errors, we assume the amplitudes of driven pulse to vary in the range of $(1 + \alpha) {\Omega _{0}}$ with the error fraction $\alpha\in[-0.1,0.1]$. Secondly, we set the frequency detuning error to be $\Delta|e\rangle\langle e|$ with $\Delta=\beta\Omega_{0}$ being static and the fraction is $\beta\in[-0.1,0.1]$. As show Figs.~\ref{fig3}, we plot the \emph{T} (\ref{fig3}a) and $X^{1/2}$ (\ref{fig3}b) gate fidelities of B-NHQC, CB-NHQC, NHQC and C-NHQC as functions of the error fraction $\alpha$ and $\beta$ with the relaxation, and find our CB-NHQC strategy is indeed the most robust than the other schemes against both Rabi errors and detuning errors, see Appendix B for the detail analysis. Thirdly, we also plot the \emph{T} and $X^{1/2}$ gate fidelities as a function of decoherence rate $\Gamma$ for all the four schemes. As shown in Fig.~\ref{fig3}(e) and \ref{fig3}(f), our schemes of B-NHQC and CB-NHQC can greatly suppress the decoherence effect comparing with NHQC and C-NHQC.

\begin{figure}[tb]\centering
\includegraphics[width=8.5cm]{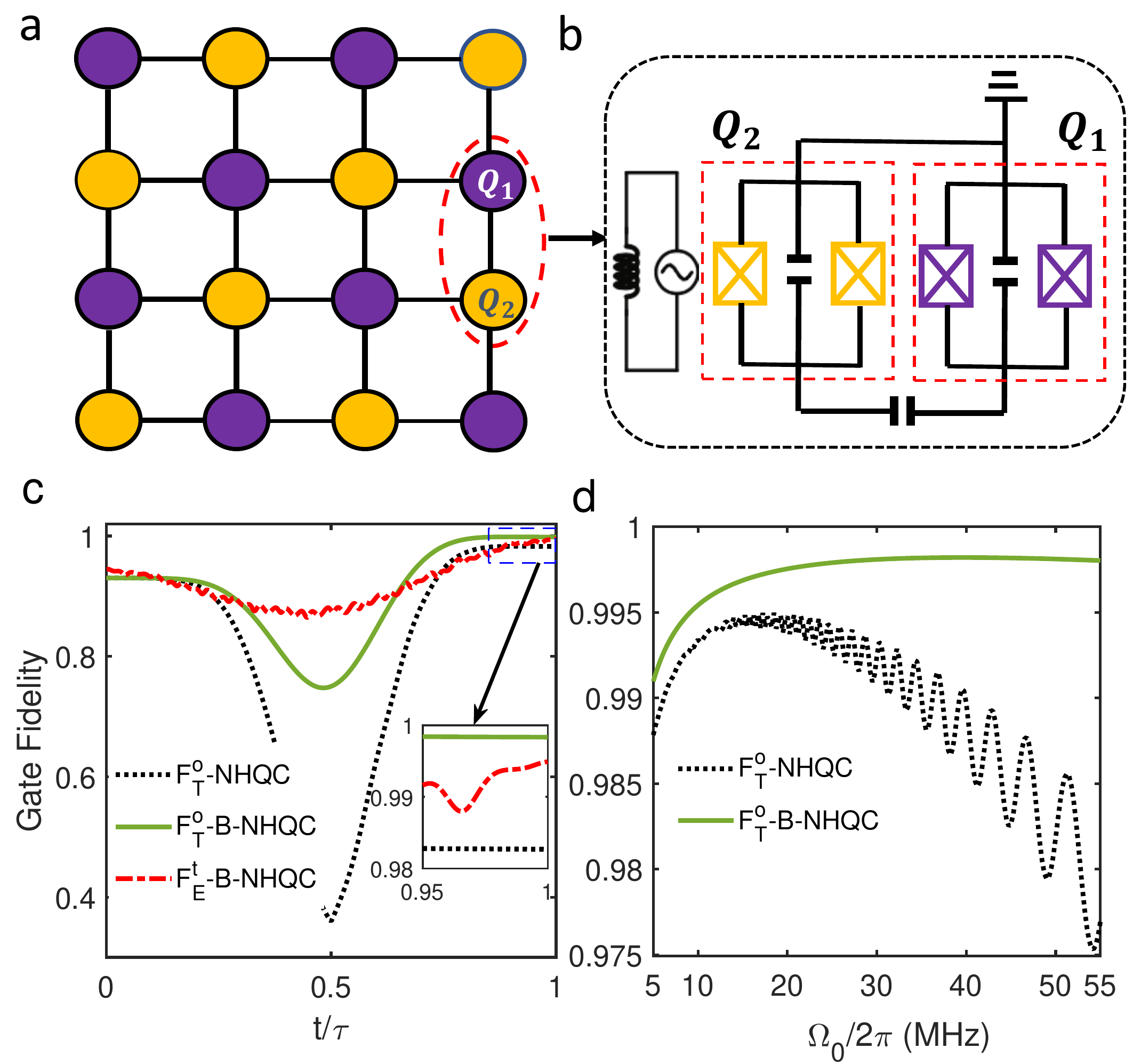}
\caption{\label{Fig4}(a) Scale-up of our physical scheme, the 2D square qubit lattice includes capacitively coupled superconducting transmon qubits, where the transmon qubits are denoted by filled circles.  (b) Schematic of our circuit consisting of two capacitively coupled qubits, where $Q_{2}$ is biased by an ac magnetic flux to periodically modulate its transition frequency.
(c) The \emph{T} and two-qubit entangled gate fidelities of B-NHQC as a function of the time. (d) The gate fidelities of the B-NHQC and NHQC geometric \emph{T} gates as functions of
the tunable parameters $\Omega_{0}$.}
\end{figure}

\emph{Physical realization on superconducting circuits.}---Here, we illustrate the implementation of our idea on superconducting circuits, as shown in Figs.~\ref{Fig4}(a) and \ref{Fig4}(b). We consider the three lowest levels of a transmon qubit, where the state $| g \rangle\equiv |0\rangle$ and the state $| f \rangle \equiv|1\rangle$ are chosen as our qubit logic states; while the state $|e\rangle$ as an auxiliary state,  the third excited state $|h\rangle\equiv|2\rangle$ as a leakage state. The corresponding Hamiltonian, in the interaction picture, is approximately given by Eq. (\ref{h1}). Thus, we can realize B-NHQC single-qubit gate in a single-loop way with a superconducting transmon device.

In our simulation, we choose a simple pulse $\Omega(t)=\Omega_{0}\sin^{2}(\pi t/\tau)$ to suppress the cross coupling and leakage to higher excited energy levels due to the intrinsic weak anharmonicity $\kappa$ of the transmon qubit. Using the parameters in current experiments \cite{Barends2014,Chen2016}, $\Omega_{0}=2\pi\times 45$ MHz, $\Gamma=2\pi\times 4$ kHz, and  $\kappa=-2\pi\times 260$ MHz, we found that the single-qubit \emph{T} gate fidelity $F^{o}_{T}$ can be significantly improved from 98.40\% to 99.84\% compared with NHQC under the same experimental conditions, and the minimum gate error can be reduced by as much as 64\%, 
as shown in Fig.~\ref{Fig4}(c) and~\ref{Fig4} (d).

Nontrivial two-qubit gates can be implemented on two capacitively coupled transmons,  denoting as $Q_{1}$ and $Q_{2}$, as shown in Fig.~\ref{Fig4}(b). The Hamiltonian can be written as $H_{\mathrm{sys}}=\sum_{k=1}^{2}H_{q_{k}}+H_{q_{c}}$ where $H_{q_{k}}$ is the single-qubit Hamiltonian and $H_{q_{c}}$ is the coupling term. With the lowest four levels of a transmon being considered,
the free single-qubit Hamiltonian is $H_{q_{k}}=\omega_{q_{k}} | e\rangle\left\langle e\left|+\left(2 \omega_{q_{k}}+\kappa_{k} \right)\right| 1\right\rangle\langle 1 |+\left(3\omega_{q_{k}}+2\kappa_{k} \right)| 2\rangle\langle 2|$,
where $\omega_{q_{k}}$ is the resonant frequency of a transmon, $\kappa_{k}$ is the corresponding anharmonicity. Meanwhile, the two-qubit coupling Hamiltonian is $H_{q_{c}}=g_{12}\left(\sigma_{1}^{\dagger} \otimes\sigma_{2}+\mathrm{H.c.}\right)$, where $g_{12}$ is the static capacitive coupling strength,
and $\sigma_{i}= |0\rangle\langle e|+\sqrt{2}| e\rangle\langle 1 |+\sqrt{3}| 1\rangle\langle 2 |$ is the lower operator for the transmon. To obtain tunable coupling between the two qubits, we add an ac magnetic flux on transmon $Q_{2}$ to periodically modulate its eigen-frequency as
$\omega_{q_{2}}(t)=\omega_{q_{2}}+\varepsilon \sin \left(\nu t\right)$,
where $\varepsilon$ and $\nu$ are the modulation amplitude and frequency, respectively. Moving into the interaction picture, the interaction Hamiltonian reads~\cite{Reagor2018,Caldwell2018,Li2018,Chen2018}
\begin{eqnarray}\label{HI}
{ H }_{ I }&=&g_{ 12 }[ { \sqrt { 2 } e }^{ i\left( \Delta _{ 1 }-\kappa _{ 2 } \right) t }{ e }^{ i\beta\cos  \left( \nu t \right)  }|ee\rangle \langle 01\notag\\
&&+\sqrt { 2 } { e }^{ i\left( \Delta _{ 1 }-\kappa _{ 1 } \right) t }{ e }^{ i\beta\cos  \left( \nu t \right)  }|10\rangle \langle ee|\notag\\ &&+\sqrt { 6 }{ e }^{ i\left( \Delta _{ 1 }-\kappa _{ 2 }+2\kappa _{ 1 } \right) t } { e }^{ i\beta \cos  \left( \nu t  \right)  }|2e\rangle \langle 11|\\
&&+\sqrt { 6 } { e }^{ i\left( \Delta _{ 1 }-2\kappa _{ 2 }+\kappa _{ 1 } \right) t }{ e }^{ i\beta \cos  \left( \nu t  \right)  }|11\rangle \langle e2|+\text{H.c.}],\notag
\end{eqnarray}
where $\Delta_{1}=\omega_{1}-\omega_{2}$, $\beta=\epsilon / \nu$, and $| m n \rangle\equiv| m \rangle_{1} \otimes | n \rangle_{2}$. From the above Hamiltonian, resonant interaction can be induced  in both the two or four excitation subspaces, by  different choice of the driving frequency $\nu$. Ignoring the higher-order oscillating terms, when $\Delta_{1}-\kappa_{2}-\mu=\nu$ with a small detuning $\mu$, we can get the effective Hamiltonian in the subspace $\left\{|01\rangle,|11\rangle,|e2\rangle,|ee\rangle \right\}$ as,
\begin{equation}\label{Eff}
H_{e}=\frac{1}{2} g_{12}^{\prime} \left[e^{-i \mu t} (| ee \rangle\langle 01 |+3e^{i(\kappa_{1}-\kappa_{2}) t} | 11 \rangle\langle e2 |)+H . c.\right],
\end{equation}
where $g^{\prime}_{12}=2 \sqrt{2} g_{12} J_{1}\left(\beta\right)$ with $J_{m}\left(\beta\right)$ being the Bessel function of the first kind.

In the same way as the single-qubit phase gate case, we can realize the two-qubit entangled gate with minimum time by solving the QBE, and we obtain that $\mu=2(\xi-\pi)/\pi$ with a minimum gate time
$\tau_{2}=2 \sqrt{\pi^{2}-(\pi-\xi)^{2}}/g^{\prime}_{12}$, where $\xi_{1}$ and $\xi_{2}$ corresponds to the obtained unconventional geometric phase in the subspaces $\left\{|01\rangle,|ee\rangle \right\}$ and $\left\{|11\rangle,|e2\rangle \right\}$. And the obtained evolution operator $U_{E}(\tau_{2})$, i.e., t
he holonomic entangling gate ($\xi_{1}\neq\xi_{2}$), in the two-qubit basis $\{ | 00\rangle, | 01 \rangle, | 10 \rangle, | 11 \rangle \}$ as
\begin{equation}
U_{E}(\xi_{1},\xi_{2})=\text{diag}\left(1,e^{i \xi_{1}},  1, e^{i \xi_{2}} \right).
\end{equation}

To  evaluate the gate performance, we take the control \emph{T} gate $U_{E}(\pi/4,-\pi/4)$ as a typical example. Here, we set the parameters ~\cite{Reagor2018,Caldwell2018} of the transmons as $\kappa_{1}=-2\pi\times 220$ MHz, $\kappa_{2}=\kappa$, $\Delta_{1}=2\pi\times 146$ MHz and $g_{12}=2\pi\times 10$ MHz.  For a general initial state $| \psi(0) \rangle=\left(\cos \chi_{1} | 0\right\rangle_{1}+\sin \chi_{1} | 1 \rangle_{1} ) \otimes\left(\cos \chi_{2} | 0\right\rangle_{2}+\sin \chi_{2} | 1 \rangle_{2} )$, the two-qbuit gate fidelity defined by~\cite{Chen2018} $F^{t}_{E}=\left(1 / 4 \pi^{2}\right) \int_{0}^{2 \pi} \int_{0}^{2 \pi} \left\langle\psi_{E}\left|\rho_{2}\right| \psi_{E}\right\rangle d \chi_{1} d \chi_{2}$ with the target state   $|{\psi _{ E } } \rangle=U_{E}|\psi(0)\rangle$, the gate fidelity of $U_{E}$ can be as high as 99.50\%, as shown in Fig. \ref{Fig4}(c).

\emph{Conclusion.}---We have proposed an unconventional approach  of  NHQC scheme with non-Abelian geometric phase, which can be compatible with time-optimal control technology to realize the fastest holonomic gate. Comparing with conventional NHQC, our proposal is more robust against the experimental control errors and decoherence. We also presented an explicit way to implement our scheme using a three-level system, and numerically simulated the performance of pulse optimization for superconducting circuits, where the gate fidelity can be significantly improved. Moreover, we discuss how the B-NHQC gate presented here can be applied to two-qubit gates in detail.

\acknowledgements
We thank T. Chen for helpful discussion. This work is supported by the Key-Area Research and Development Program of Guangdong Province (Grant No. 2018B030326001), the National Natural Science Foundation of China (Grant No. 11875160 and No. 11874156),  the Natural Science Foundation of Guangdong Province (Grant No. 2017B030308003), the National Key R\& D Program of China (Grant No. 2016YFA0301803, the Guangdong Innovative and Entrepreneurial Research Team Program (Grant No. 2016ZT06D348), the Economy, Trade and Information Commission of Shenzhen Municipality (Grant No. 201901161512), the Science, Technology and Innovation Commission of Shenzhen Municipality (Grant No. JCYJ20170412152620376, No. JCYJ20170817105046702, and No. KYTDPT20181011104202253).

\section{Appendix A: Geometric nature of the unconventional holonomic gate}

To verify the Eq. (\ref{UHOLONOMY})  is a holonomic gate, i.e., we need the conditions of gauge invariance and parallel transport being satisfied in geometry. Here, we first check the the condition of gauge invariance. A gauge transformation is defined to link two frameworks~\cite{b3,b4}, e.g., $\left|\phi_{a}\right\rangle \rightarrow \sum_{l}\left|\phi_{l}\right\rangle R_{l a}\left[\lambda_{a}(t),\eta_{b}(t) \right]$ with $R$ being the unitary transformation operator such that $R(0)=R\left[\lambda_{a}(\tau),\eta_{b}(\tau) \right]$, then $A\rightarrow R^{\dagger} A R+i R^{\dagger} \dot{R}$ and $K\rightarrow R^{\dagger} K R$. Consequently, we know that $A+K=A^{\lambda}\rightarrow R^{\dagger} A^{\lambda} R+i R^{\dagger} \dot{R}$ also satisfies the gauge transform. Note that the unconventional holonomy $U(\tau)$ transforms as $U(\tau) \rightarrow R^{\dagger}(0) U(\tau) R(0)$, and thus it also shares the gauge covariance. For the special example, under a gauge transformation, i.e., $\left|\phi_{a}\left[\lambda_{a}(t), \eta_{b}(t)\right]\right\rangle \rightarrow e^{i \alpha_{a}(t)}\left|\phi_{a}\left[\lambda_{a}(t), \eta_{b}(t)\right]\right\rangle$, then the gauge potential $A^{\lambda}_{aa}=\dot{\lambda}_{a}$ transforms
\begin{equation}\label{GAUTS}
    \dot{\lambda}_{a} \rightarrow \dot{\lambda}_{a}+\frac{d \alpha_{a}}{d t} \ .
\end{equation}
and this is similar to the vector potential in electrodynamics.

To continue, we consider the condition of parallel transport condition in the following. The state $|\Psi(t)\rangle$ is parallel transported if the following condition~\cite{Simon1983}
\begin{equation}\label{PARA}
\langle\Psi(t)|\frac{d}{dt}|\Psi(t)\rangle=0 \ ,
\end{equation}
is satisfied.  Here, we follow the pioneer works in Ref.~\cite{Simon1983,Samuel1988} to verify the condition of parallel transport in our work. Let
us define a new state vector $|\tilde{\phi}_{a}\left(\lambda_{a}, \eta_{b}\right)\rangle=e^{i\int h_{a}(t) dt}\left|\psi_{a}\left(\lambda_{a}, \eta_{b}\right)\right\rangle$ to remove the dynamical phase with $h_{a}(t)=\left\langle\psi_{a}(\lambda_{a}, \eta_{b})|H| \psi_{a}(\lambda_{a}, \eta_{b})\right\rangle$, where $\left|\psi_{a}\left(\lambda_{a}, \eta_{b}\right)\right\rangle=e^{i \lambda_{a}}\left|\psi_{a}\left(\eta_{b}\right)\right\rangle$ satisfies the Schr\"{o}dinger equation. Explicitly,  the condition of parallel transport is satisfied, i.e.,
\begin{eqnarray}\label{PROVE}
&&\langle\tilde{\phi}_{a}\left(\lambda_{a}, \eta_{b}\right)|i \frac{d}{d t}| \tilde{\phi}_{a}\left(\lambda_{a}, \eta_{b}\right)\rangle\notag\\
&=&-h_{a}+\left\langle\phi_{a}\left(\lambda_{a}, \eta_{b}\right)|H| \phi_{a}\left(\lambda_{a}, \eta_{b}\right)\right\rangle=0 \ ,
\end{eqnarray}
Consequently, Eq.(\ref{PROVE}) implies
\begin{equation}\label{RR1}
\dot{\lambda}_{a}+i\langle\tilde{\phi}_{a}\left(\lambda_{a}, \eta_{b}\right)|\frac{\partial}{\partial \eta_{b}}| \tilde{\phi}_{a}\left(\lambda_{a}, \eta_{b}\right)\rangle \frac{d \eta_{b}}{d t}=0 \ ,
\end{equation}
hence, we obtain the purely geometric phase $\lambda_{a}$ by integrating over a closed circuit as
\begin{equation}\label{geometricPhase}
\lambda_{a}=-i \oint\langle\psi_{a}\left(\eta_{b}\right)|\frac{d}{d \eta_{b}}| \psi_{a}\left(\eta_{b}\right)\rangle d \eta_{b} \ .
\end{equation}
The geometric phase $\lambda_{a}$ depends only on the global geometric feature of the evolution path.


\section{Appendix B: Theoretical Robustness analysis of B-NHQC}

Here, we theoretically investigate the robustness of B-NHQC against the inevitable experimental errors, i.e., Rabi control and detuning errors, with small error fractions of $\alpha$ and $\beta$, respectively. In other words, the actually Hamiltonian becomes $H(t)\rightarrow H(t)+H_{\alpha}(t)+H_{\beta}(t)$, where $H_{\alpha}(t)=\alpha H(t)$ and $H_{\beta}(t)=\beta V=\beta\Omega_{0}|e\rangle \langle e|$ are the perturbing Hamiltonian.

\subsection{A. Robustness against Rabi control error}

We firstly consider the B-NHQC gate under the Rabi control error with a small error fraction $|\alpha| \ll 1$ while $\beta=0$. By using perturbation theory, the orthogonal solution in Eq. (\ref{h1}) with the error at time $t=\tau$ becomes,
\begin{equation}\label{SA1}
\left|\psi_{m}^{\prime}(\tau)\right\rangle=N_{m}\left(\left|\psi_{m}(\tau)\right\rangle-i \alpha \sum_{k=0}^{2} Q_{k m}\left|\psi_{k}(\tau)\right\rangle\right) \ ,
\end{equation}
where $m=0,1,2$ and $Q_{k m}\equiv \int_{0}^{\tau}\left\langle\psi_{k}(t)|H(t)|\psi_{m}(t)\right\rangle d t$. The term $N_{m}=\left(1+\alpha^{2} \sum_{k=0}^{2}\left|Q_{k m}\right|^{2}\right)^{-1 / 2}$ is the state normalized coefficient.
In this way, the evolution operator under control noise is given by
\begin{widetext}
\begin{eqnarray}\label{SA2}
U_{E}^{\prime}(\tau) =\sum_{m=0}^{2}\left|\psi_{m}^{\prime}(\tau)\right\rangle\left\langle\psi_{m}(0)\right| =\sum_{m=0}^{2}N_{m}\left(|\psi_{m}(\tau)\rangle\langle\psi_{m}(0)|-i \alpha \sum_{k=0}^{2} Q_{k m}| \psi_{k}(\tau)\rangle\langle\psi_{m}(0)|\right) \ .
\end{eqnarray}
Note that the evolution operator $ U_{E}^{\prime}(\tau)$ can be spanned in the logical $\{|0\rangle,|1\rangle\}$ basis, i.e.,
\begin{equation}\label{SA3}
 U^{\prime}(\tau)=\sum_{l,n=0,1}\left[\langle l|U_{E}^{\prime}(\tau)|n\rangle\right]|l\rangle \langle n| \ .
\end{equation}
To evaluate the performance of the gate caused by the Rabi control error, the gate fidelity~\cite{Souza2011,Genov2017} is taken by
\begin{eqnarray}\label{SA4}
F=\frac{1}{2}\left|\operatorname{Tr}\left[U^{\prime}(\tau) U^{+}(\tau)\right]\right|\approx \left|\frac{1}{2}+\sum_{l=0}^{1} \frac{N_{l}}{4 N_{1} N_{2}}\left\{\left(1-i \alpha Q_{l l}\right)\left[1+(-1)^{l+1} \cos \eta_{3}\right]-i \alpha Q_{(1-l) l} \sin \eta_{3}\right\}\right| \ ,
\end{eqnarray}
where $U(\tau)$ is the ideal B-NHQC gate in Eq. (\ref{BUNHQC}). Using  Eqs. (\ref{h1}) and  (\ref{ThreST}) , the gate fidelity in Eq. (\ref{SA4}) can be simplified as
\begin{equation}\label{SA5}
F=\sqrt{\left(\frac{N_{1}+1}{2 N_{1}}\right)^{2}+\frac{\alpha^{2}}{4 N_{1}^{2}} \sin ^{2}\left[\eta_{3}+\tan ^{-1}\left(\frac{\left|Q_{00}\right|}{\operatorname{Re}\left(Q_{01}\right)}\right)\right]} \ ,
\end{equation}
where $Q_{00}=\frac{1}{2} \int_{0}^{\tau} \dot{\eta}_{2} (1/\cos\eta_{3}-\cos \eta_{3}) dt$, $Q_{01}=\frac{1}{2} \int_{0}^{\tau} e^{i\int_{0}^{t}\frac{\dot{\eta}_{2}}{\cos\eta_{3}}dt^{\prime}}(\dot{\eta}_{2} \sin \eta_{3}) dt$ and $Q_{02}=0$. Therefore, we can theoretically evaluate the performance of different B-NHQC gates with different settings of $\eta_{2}(t)$ and $\eta_{3}$. For example, we obtain $Q_{00}=-3\pi/4$ and $Q_{01}=0$ with $\eta_{3}=2\pi/3$ and $\eta_{2}(t)=-\pi t/\tau$ for the $X^{1/2}$ gate of B-NHQC, which reduces to the $X^{1/2}$ gate in NHQC~\cite{Sjoqvist2012,Xu2012} by setting of $\eta_{3}=\pi/2$ and $\eta_{2}=0$ with $t\in[0,\tau/2]$ and $\eta_{2}=-\pi/2$ with $t\in[\tau/2,\tau]$. Consequently, we can also evaluate the robustness of NHQC $X^{1/2}$ gate with the corresponding $Q_{00}=0$ and $Q_{01}=i\pi\sin(\pi/4)$ by using Eq. (\ref{SA5}). As shown in Fig.~\ref{figs1}(a), we plot the $X^{1/2}$ gate fidelities of B-NHQC and NHQC as a function of the  error fraction $\alpha$, where we find that  both the B-NHQC and NHQC gates share similar performance in terms of the Rabi control error. Moreover, we can clearly see that the numerical results are in good agreement with the theoretical result predicted in Eq. (\ref{SA5}).

\begin{figure}[tb]
	\centering
\includegraphics[width=8.5cm]{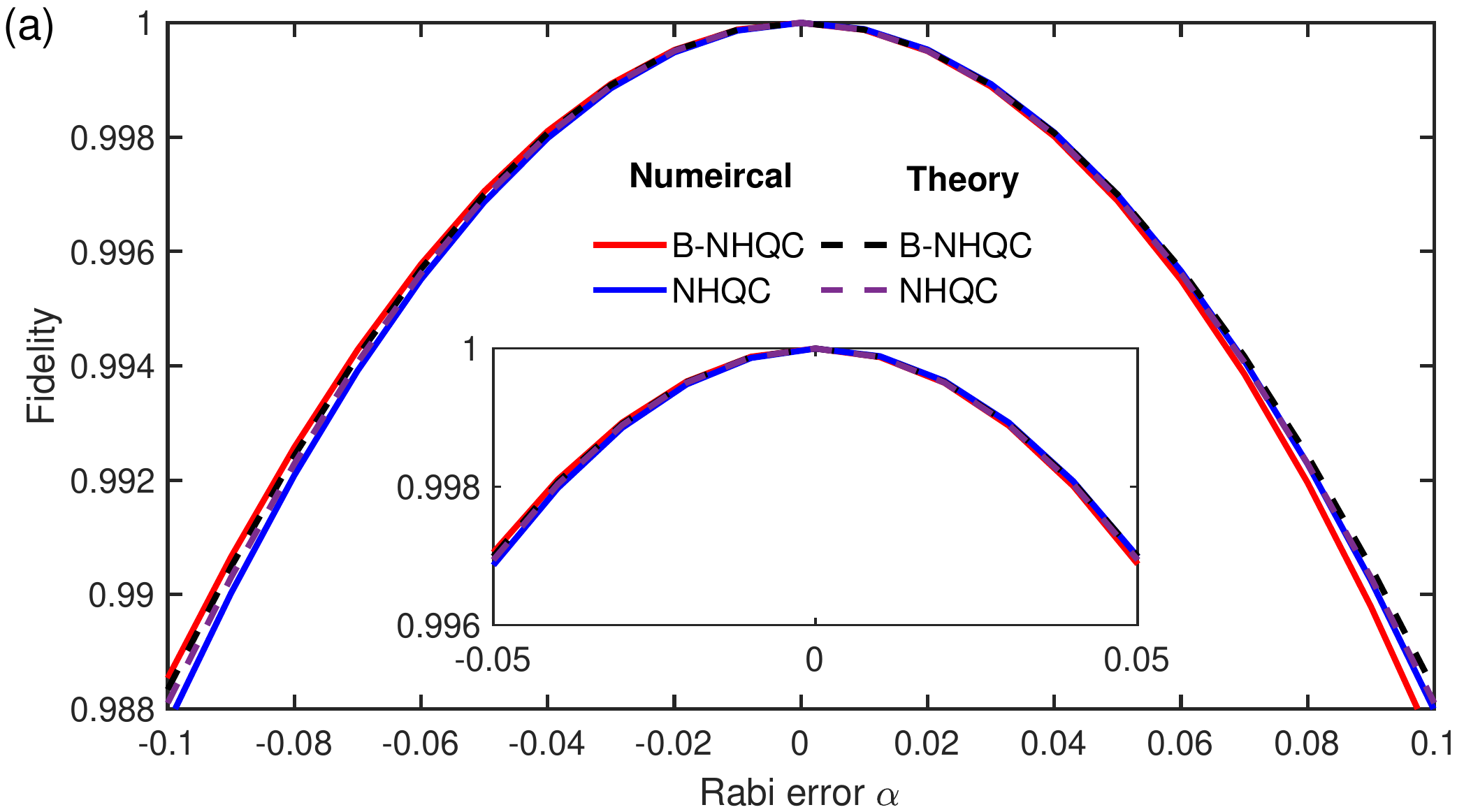}
\includegraphics[width=8.5cm]{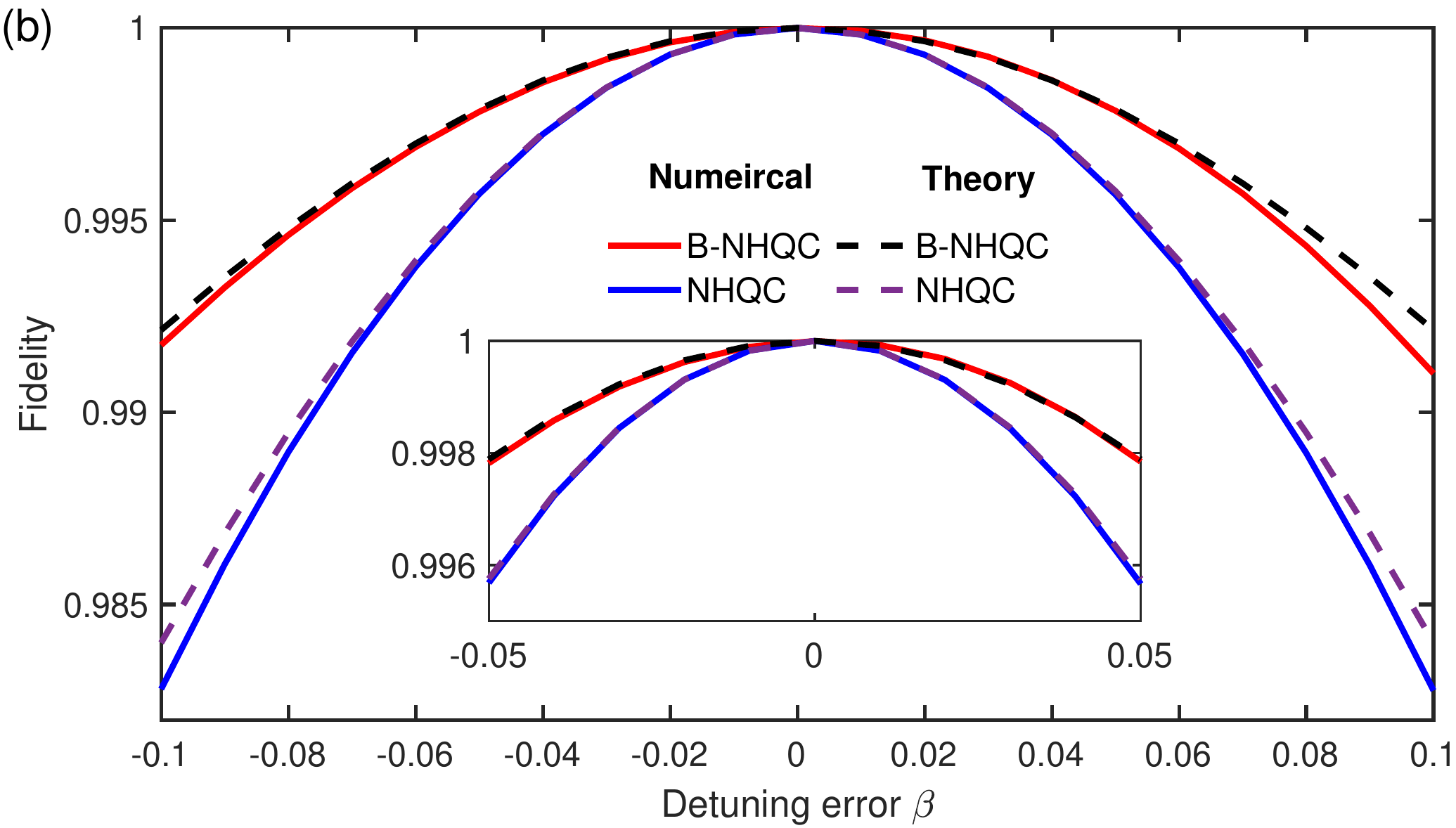}
\caption{\label{figs1} The gate fidelity under imperfections without the consideration of decoherence. The $X^{1/2}$ gate fidelity for B-NHQC and NHQC cases under (a) the Rabi error, and (b) detuning error, respectively.
The solid curve are numerical results and the dashed curve is the theory.}
\end{figure}

\subsection{B. Robustness against detuning error}
To evaluate the gate robustness of B-NHQC against the detuning error, we set the error fractions as $|\beta| \ll 1$ and $\alpha=0$. Similar to the above theoretical robustness analysis under the Rabi
control error, we can also obtain the gate fidelity with the detuning error as
\begin{equation}\label{SA6}
F\approx \left|\frac{1}{2}+\sum_{l=0}^{1} \frac{W_{l}}{4 W_{1} W_{2}}\left\{\left(1-i \beta P_{l l}\right)\left[1+(-1)^{l+1} \cos \eta_{3}\right]-i \beta P_{(1-l) l} \sin \eta_{3}\right\}\right| \ ,
\end{equation}
\end{widetext}
where $P_{k m}\equiv \int_{0}^{\tau}\left\langle\psi_{k}(t)|V|\psi_{m}(t)\right\rangle d t$, $W_{11}=\sqrt{1+\beta^{2}\left(\left|P_{00}\right|^{2}+\left|P_{10}\right|^{2}\right)}$ and $W_{22}=\sqrt{1+\beta^{2}\left(\left|P_{11}\right|^{2}+\left|P_{01}\right|^{2}\right)}$. With the help of the Eq. (\ref{h1}) and Eq. (\ref{ThreST}), we get $P_{00}=\int_{0}^{\tau} \Omega_{0} \cos ^{2} \frac{\eta_{3}}{2} d t$, $P_{11}=\int_{0}^{\tau} \Omega_{0} \sin ^{2} \frac{\eta_{3}}{2} d t$, $P_{01}=P_{10}^{*}=-\int_{0}^{\tau} \frac{\Omega_{0}}{2} \sin \eta_{3} e^{i \int_{0}^{t} \frac{\dot{\eta}_{2}}{\cos \eta_{3}} d t^{\prime}} d t$.  As shown in Fig.~\ref{figs1}(b), we also plot the $X^{1/2}$ gate fidelities of B-NHQC and NHQC as a function of the detuning error fraction $\beta$, where both the numerical and theoretical results show that B-NHQC can greatly suppress the detuning error comparing with NHQC~\cite{Sjoqvist2012,Xu2012}.


\end{document}